\documentclass{epl}
\usepackage{epsfig}
\title{Re-entrant pinning of Wigner molecules in a magnetic field due to a Coulomb impurity}
\author{B. Szafran\inst{1,2} \and F.M. Peeters\inst{1}}
\institute{
  \inst{1} Departement Natuurkunde, Universiteit Antwerpen
(Campus Drie Eiken), B-2610 Antwerpen, Belgium \\
  \inst{2} Faculty of Physics and Nuclear Techniques, AGH
University of Science and Technology, al. Mickiewicza 30, 30-059
Krak\'ow, Poland }

\shorttitle{Re-entrant pinning of Wigner molecules}
\begin{document}

\maketitle

\begin{abstract}
Pinning of magnetic-field induced Wigner molecules (WMs) confined
in parabolic two-dimensional quantum dots by a charged defect is
studied by an exact diagonalization approach. We found a
re-entrant pinning of the WMs as function of the magnetic field, a
magnetic field induced re-orientation of the WMs and a
qualitatively different pinning behaviour in the presence of a
positive and negative Coulomb impurity.
\end{abstract}

Low-density electron systems in bulk may form an ordered
crystalline phase called Wigner crystal~\cite{WC} in which
electron charges are spatially separated. A similar collective
type of electron localization in quantum dots (QDs) is called
Wigner molecule (WM)~\cite{Egger}. WMs may be formed in large
QDs~\cite{Egger} or be induced by a strong magnetic
field~\cite{Koonin} in the quantum Hall regime. Wigner
localization is observed in the inner coordinates of the quantum
system whose charge density conserves the symmetry of the external
potential~\cite{REIMAN}. Therefore, in circular
QDs~\cite{REIMAN,MAKSYM} the charge density will be circular
symmetric even in the Wigner phase. However, a perturbation of the
potential may pin~\cite{pinedWC} the charge density at a fixed
orientation in the laboratory frame which should allow for the
experimental observation~\cite{CDM} of Wigner localization.
Pinning of the magnetic-field induced WMs by the anisotropy of the
potential~\cite{Maniani} or by an attractive Gaussian impurity
potential~\cite{REGGER} in the absence of a magnetic field have
been studied previously. Here, we will show that the WM pinning is
qualitatively very different in the presence of a positive and
negative impurity.

We consider WMs induced by a magnetic field in a two-dimensional
harmonic QD. A strong magnetic field polarizes the spins of the
confined electrons and leads to the formation of a so-called
maximum density droplet (MDD) corresponding to the lowest Landau
level filling factor $\nu=1$. Stronger fields induce the MDD to
decay into a molecular phase with $\nu<1$, for which the
distribution of electrons in the inner coordinates resembles the
equilibrium configuration of a classical point-charge
system~\cite{Bedanov}. The external magnetic field increases the
absolute value of the angular momentum of the confined electron
system inducing its changes between certain 'magic'~\cite{magic}
values for which the classical distribution of electrons in the
inner ('rotating') frame of reference can be realized.

In this letter we investigate the way in which the potential of a
charged defect (donor or acceptor ion) situated outside the QD
symmetry axis stops the ''rotation'' of the electron system and
freezes the WM at a fixed orientation. We use the configuration
interaction approach which allows for an exact solution of the
few-electron Schr\"odinger equation. We found that at magnetic
fields inducing the angular momentum transitions the exact
ground-state can correspond to broken-symmetry charge density with
semi-classical localization in the laboratory frame.
Broken-symmetry charge distributions were previously obtained as
artifacts of mean-field methods~\cite{REIMAN}. The existance of
exact broken-symmetry states makes the WMs susceptible to pinning
by an arbitrarily distant charge defect (donor or acceptor ion) at
the angular momentum transitions. Consequently, a distant defect
induces re-entrant WM pinning as function of the strength of the
magnetic field. We show that the orientation of the pinned WMs can
change with the magnetic field and demonstrate an essentially
different pinning behavior for a positive and negative impurity.

We assume that the system of $N$-electrons is spin-polarized by
the external magnetic field and that the electrons are confined to
move in the $z=0$ plane. The present configuration interaction
approach is constructed in the following way. The single-electron
Hamiltonian for the considered system reads
\begin{equation}
h=(-i\hbar\nabla+e\mathbf{A})^2/2m^*
+m^*\omega^2(x^2+y^2)/2+Bs_zg^*\mu_B\pm e^2/4\pi\epsilon\epsilon_0
r_{ed}, \label{H1}
\end{equation}
where $m^*$ is the electron band effective mass, $\hbar \omega$ is
the confinement potential energy, $\epsilon_0$ is the static
dielectric constant, $(0,0,B)$ is the magnetic field vector, $s_z$
is the $z$-component of the electron spin, $g^*$ is the effective
Land\'e factor and $r_{ed}$ is the distance between the electron
and the charged defect. The sign in the last term of eq. (1) is
$-$ ($+$) for a positively (negatively) charged defect. We apply
the Landau gauge $\mathbf{A}=(-By,0,0)$ and adopt GaAs material
parameters $m^*=0.067m_0$, $\epsilon=12.9$ and $g^*=-0.44$ as well
as $\hbar \omega=3$ meV. Hamiltonian (1) is diagonalized in a
multicenter basis $ \Psi_\mu(\mathbf{r}) = \sum_{i=1}^{M} c_i^\mu
\psi_{\mathbf{R}_i} (\mathbf{r}) $ with
\begin{equation}
\psi_{{\mathbf R}}({\mathbf r})= \sqrt{\alpha} \exp\{-\alpha
({\mathbf r}-\mathbf{R})^2/4 +ieB(x-X)(y+Y)/2\hbar\}/\sqrt{2\pi},
\label{wv}
\end{equation}
where $\mathbf{R}=(X,Y)$. The single electron wave functions
$\Psi_\mu$ are subsequently used for construction of $M!/N!(M-N)!$
Slater determinants -- the basis set for diagonalization of the
$N$-electron Hamiltonian. $\alpha$ and the positions of the
centers $\mathbf{R}_i$ are chosen such that they minimize the
total energy. Function (2) with $\alpha=eB/\hbar$ is the lowest
Landau level eigenfunction. The basis set of displaced functions
(\ref{wv}) allows for a very precise determination of the exact
Fock-Darwin~\cite{REIMAN} energy levels, including higher
Fock-Darwin bands, which at strong magnetic fields tend to excited
Landau levels. We have verified the accuracy of the present
approach comparing its results with the standard exact
diagonalization method~\cite{sed}. We have taken 12 centers placed
on a circle. Above the MDD decay ($B>$ 5.8, 4.85 and 4.65 T for 2,
3 and 4 electrons) and below 20 T the overestimation of the exact
energy for 2, 3 and 4 electrons is lower than 0.01, 0.06 and 0.12
meV respectively. Few-electron wave functions calculated in the
Landau gauge are not eigenfunctions of the angular momentum, but
using the gauge-independent expectation value of its operator we
can look at the angular-momentum transformations of the confined
system. The precision in the determination of the critical fields
inducing ground-state transformation is better than 0.15 T.
Previously, displaced Landau level functions (2) were used in the
investigation of the WMs with approximate approaches, i.e.,
single-determinant of non-orthogonal wave functions ~\cite{Kainz},
Hartree-Fock~\cite{MCHF}, and rotated-electron-molecule
approach~\cite{Yannouleas}. Due to the arbitrariness in the choice
of centers the present configuration interaction approach can be
easily applied to potentials without circular symmetry. In the
calculations for the perturbed QDs we used 12 centers placed on an
ellipse, the size and its center of gravity were optimized
variationally.

\begin{figure}[htbp]
                \twoimages[width=6.5cm]{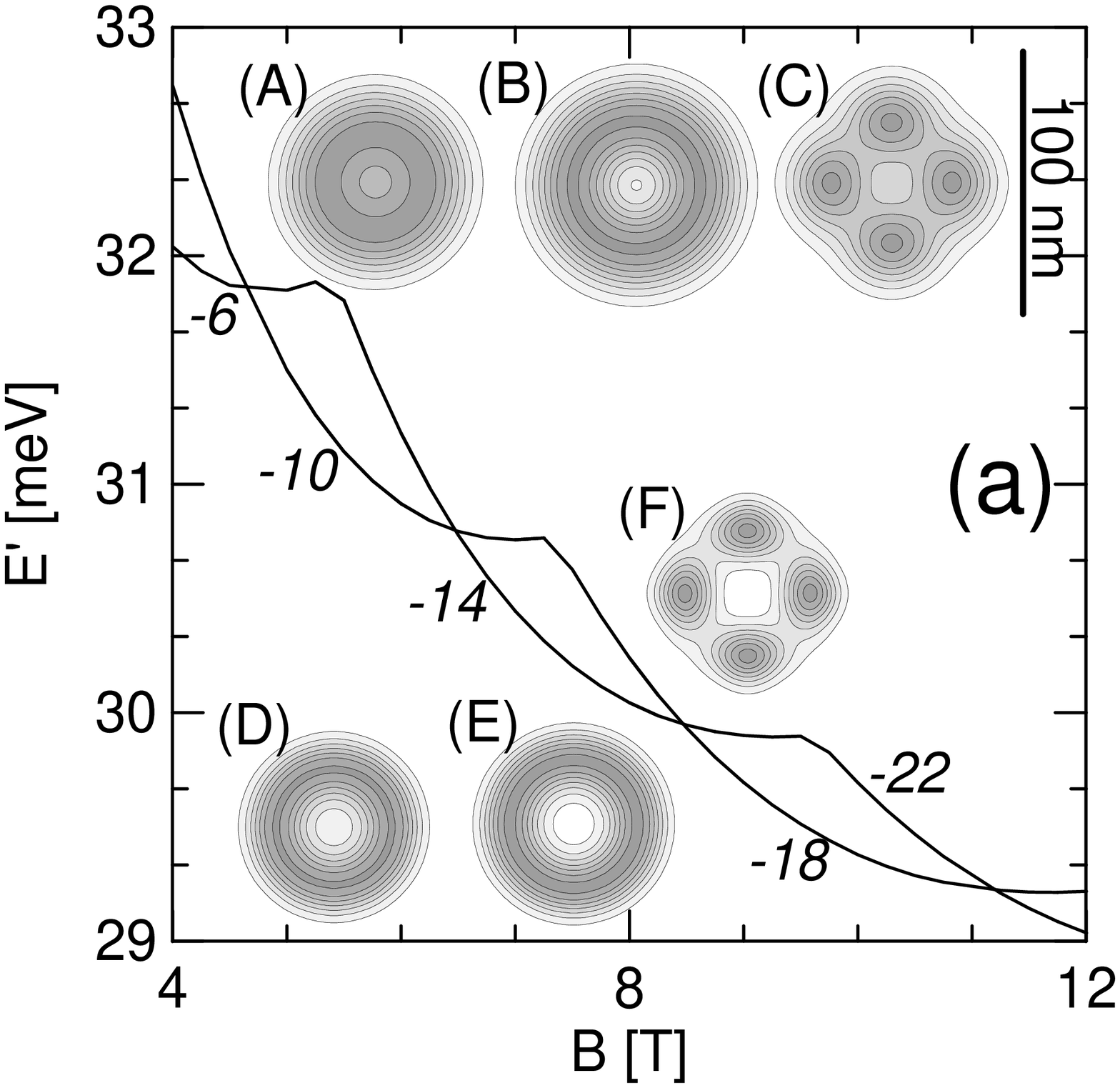}{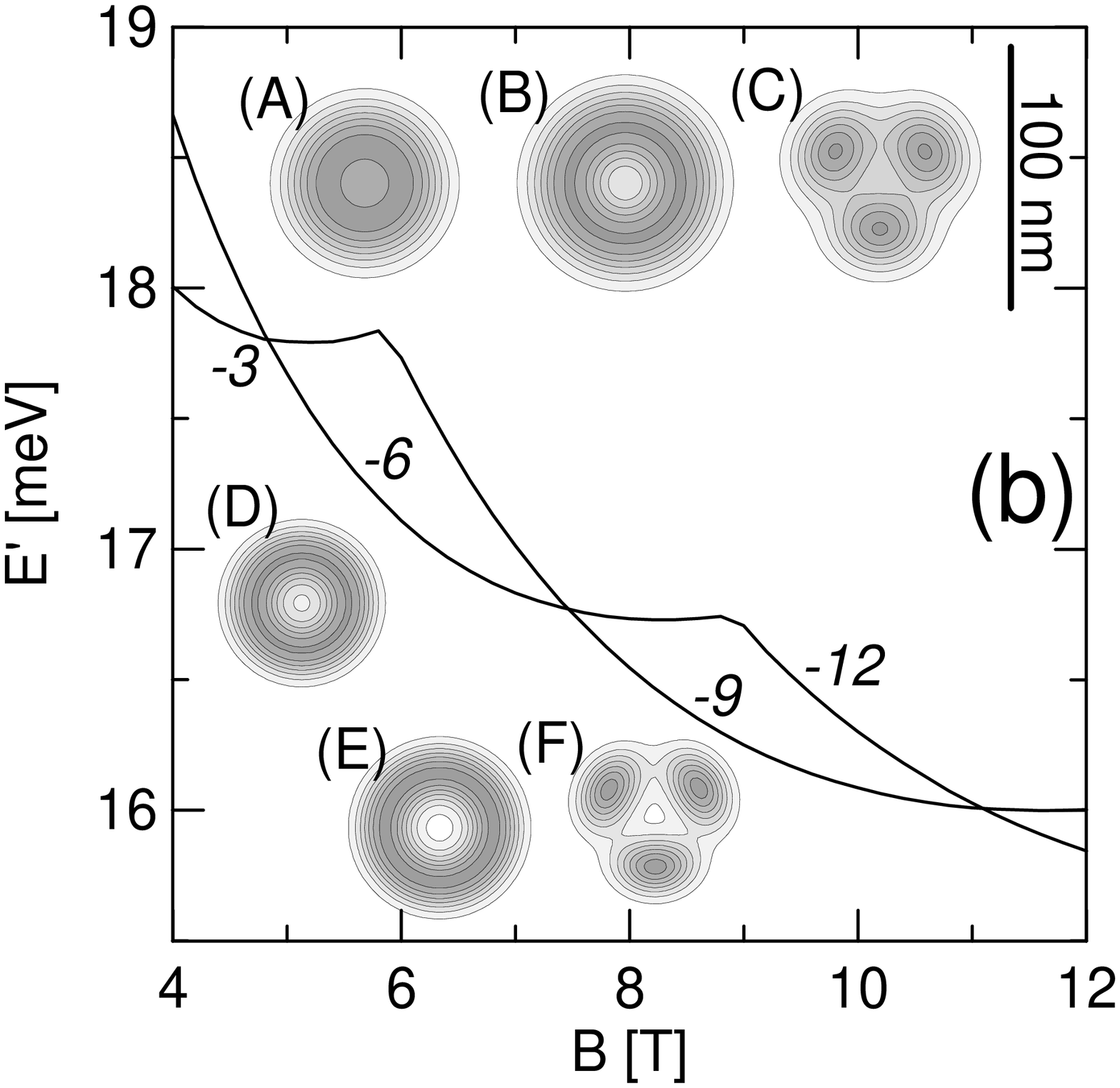}
\caption{(a) Two lowest energy levels of the 4-electron
unperturbed circular QD calculated with respect to the lowest
Landau level. Numbers denote $L$ - angular momentum in $\hbar$
units. Insets (A), (B) and (C) display the charge densities of the
states for $B=4.6$ T corresponding to $L=$ -6, -10 and broken
symmetry, respectively (the darker the shade of grey the larger
the density). (D), (E) and (F) show the charge density of the
degenerate states with $L=$ -18, -22 and broken symmetry for
$B=11.2$ T. (b) Same as (a) but for 3 electrons. (A) (B) and (C)
plotted for $B=4.2$ T correspond to $L=$ -3, -6 and broken
symmetry, respectively. (D), (E) and (F) show the charge density
of degenerate states with $L=$ -9, -12 and broken symmetry for
$B=11.2$ T.
 }
 \label{fig1}
\end{figure}

Fig. \ref{fig1} shows the two lowest 4- (a) and 3- (b) electron
energy levels calculated with respect to the lowest Landau level
($E'=E-N\times0.85$ [meV/T]) as functions of the magnetic field.
$E'$ at high field tends to the potential energy of a classical
point charge system~\cite{Kainz}. At lower magnetic fields the
ground state is the MDD with angular momentum $-N(N-1)\hbar/2$. At
larger magnetic fields the angular momentum decreases by
$N\hbar$~\cite{REIMAN,MAKSYM,magic}. The ground-state charge
density after the MDD decay has a ring-like shape with a
pronounced minimum at the center of the dot. At each ground-state
transformation the central local minimum becomes wider and the
size of the charge puddle exhibits a stepwise increase. Between
the ground state transformations the magnetic field compresses the
charge density which shrinks in a continuous fashion~\cite{sed}.

At the angular momentum transformations the ground-state charge
density is twofold degenerate. Consequently, each linear
combination of the degenerate ground-states $\Phi_1$ and $\Phi_2$
is also an eigenstate. Consider the following combination
$\Phi_{bs}=(\Phi_1+c\Phi_2)/\sqrt{2}, \label{trzy}$ with
$|c|^2=1$. Since the angular momenta of degenerate ground states
differ by $N\hbar$ the angular momentum in state $\Phi_{bs}$ is
not defined and $\Phi_{bs}$ possesses a broken-symmetry charge
distribution [cf. insets (C) and (F) in fig. \ref{fig1}]. The
charge density of the {\it exact} broken-symmetry states resembles
the {\it approximate} mean-field broken-symmetry
solutions\cite{REIMAN}. The broken-symmetry charge distributions
at high field tend\cite{MCHF} to the classical lowest-energy
distribution of point charges~\cite{Bedanov}. Superposition
$\Phi_{bs}$ extracts the inner symmetry of the magic angular
momenta states into the laboratory frame of reference. The
broken-symmetry charge distribution can be oriented at an
arbitrary angle depending on the phase of $c$.

\begin{figure}[htbp]
                \twoimages[width=6.5cm]{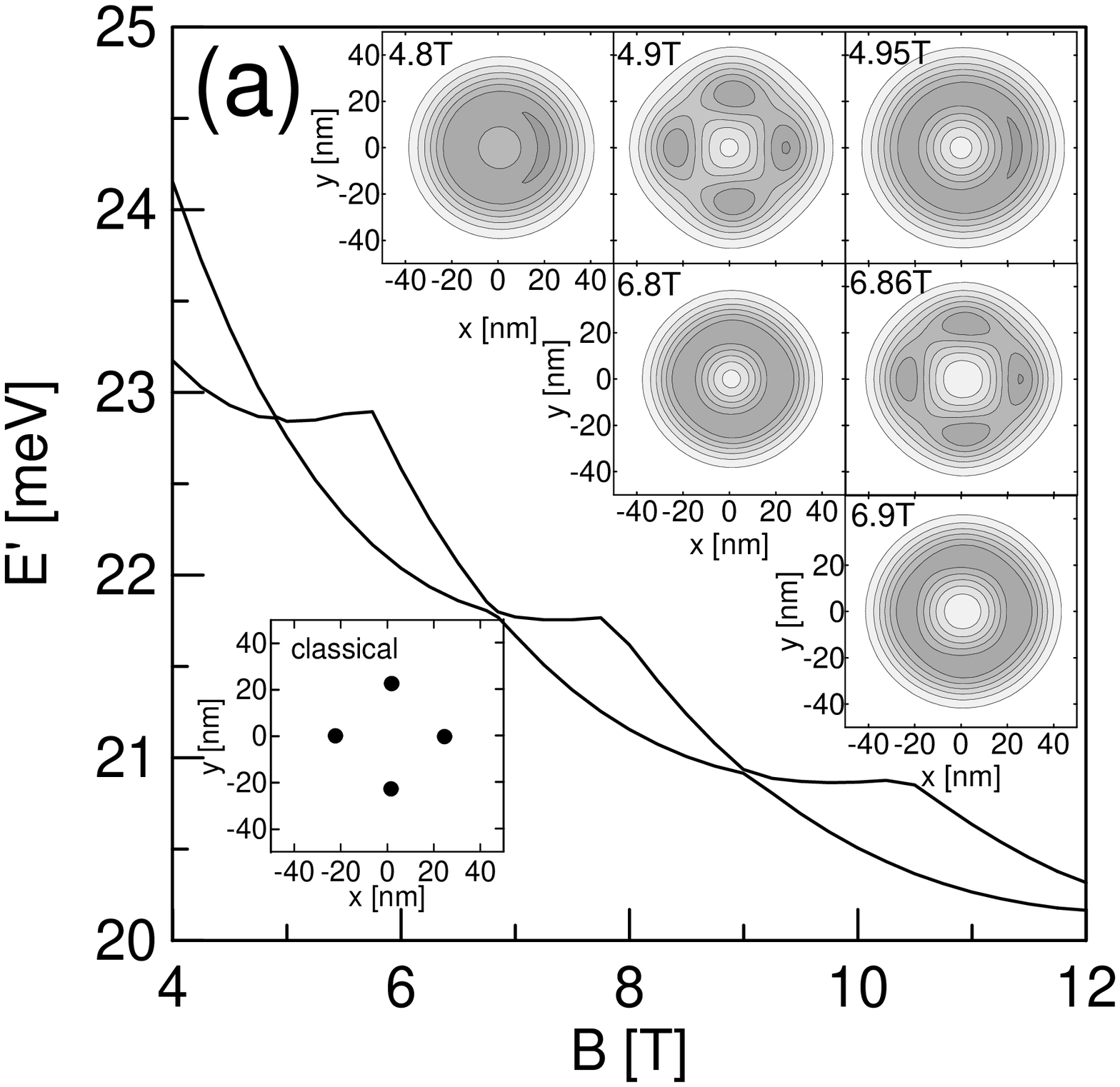}{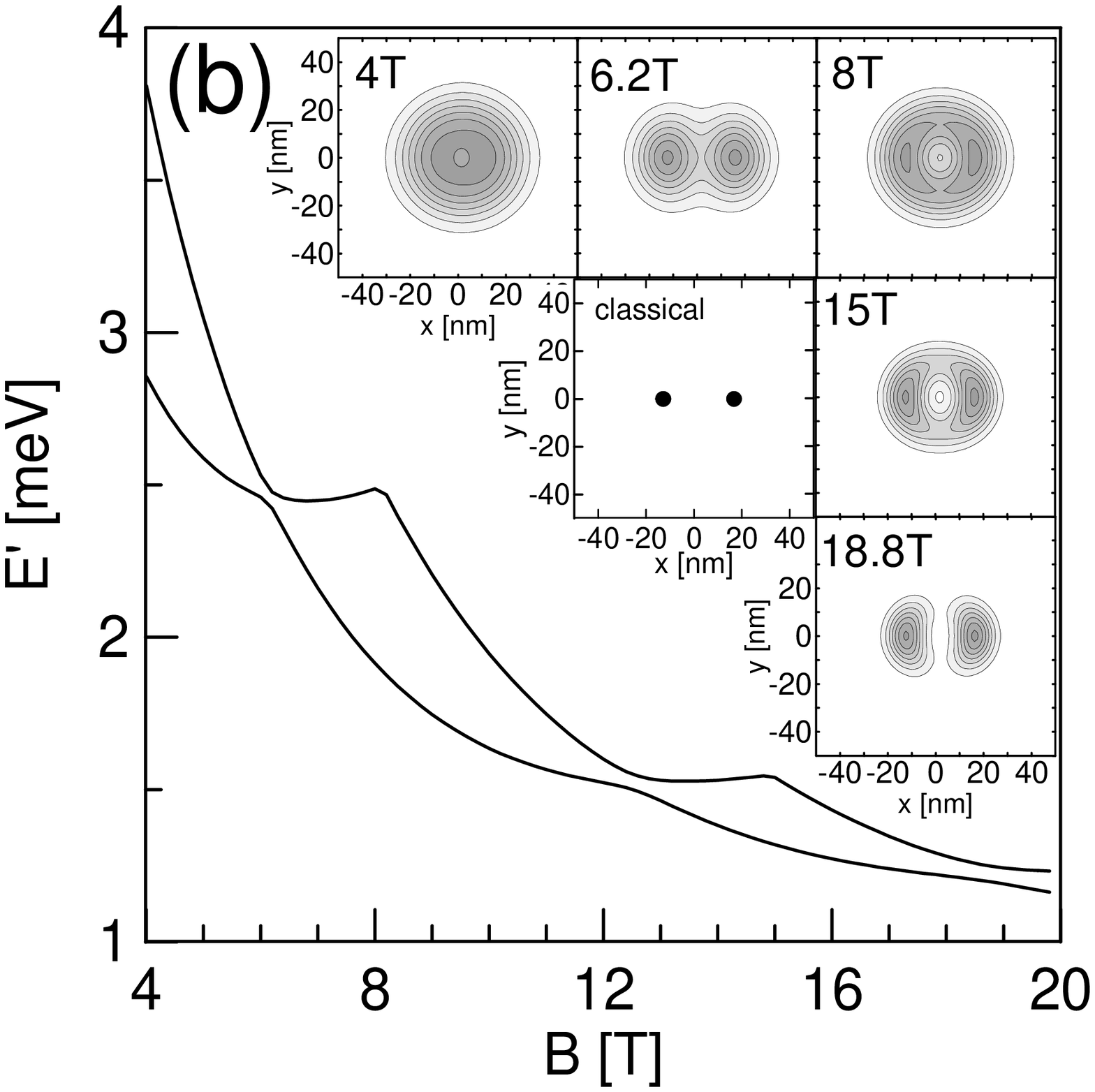}
\caption{(a) Two lowest-energy levels of the 4-electron system in
a circular QD perturbed by a potential of a positively charged
defect situated at (20,0,40) nm. Insets show the ground state
charge densities and the lowest-energy configuration of the
classical system. (b) Same as (a) but for 2-electrons.
 }
 \label{fig2}
\end{figure}

Let us now suppose that at a certain distance of the quantum dot
plane there is an impurity ion located off the symmetry axis of
the dot. In vertical quantum dots\cite{VQD2} for which the
harmonic approximation of the potential is justified~\cite{BSA},
and in which the MDD decay has been observed~\cite{VQD2},
ionized~\cite{BSA} donor impurities are present at a distance of
20-30 nm from the QD plane. The defect potential perturbs the QD
circular symmetry and mixes the angular momentum eigenstates.
Level crossings are replaced by avoided crossings. Fig. \ref{fig2}
(a) shows the two lowest-energy levels and the ground state charge
density for 4 electrons with a positively charged defect situated
at point $x=20$, $y=0$, $z=40$ nm. The energy gaps in the avoided
crossings are very small ($\sim 10^{-3}$ meV). At the avoided
level crossings (see insets for $B=4.9$ and 6.86 T)  Wigner
crystallization in the laboratory-frame (i.e. WM pinning) can be
observed. The positions of the pinned charge density maxima
coincides with the position of classical electrons in the lowest
energy configurations [cf. lowest inset of fig. \ref{fig2}(a)].
The charge density plots for the magnetic fields outside the
avoided level crossings resembles the unperturbed circular
densities [cf. fig. 1(a)], although an increased density at the
right end of the charge puddle is visible. Since the 'momentary'
pinning is a consequence of the existence of the exact
broken-symmetry states it appears for an arbitrarily far situated
defect.

Fig. 2(b) shows that the effect of the defect on the 2-electron
spectrum and the charge density is much stronger [energy gaps are
about $5\times10^{-2}$ meV]. The oscillatory character of the
pinning as function of the magnetic field is visible. At avoided
level crossings separation of the electron charges is particularly
pronounced (see insets for 6.2 and 18.8 T). The effect of the
negatively charged defect at this rather large distance from the
QD is similar, although the molecules become pinned at different
angles.

\begin{figure}[htbp]
                \twoimages[width=6.5cm]{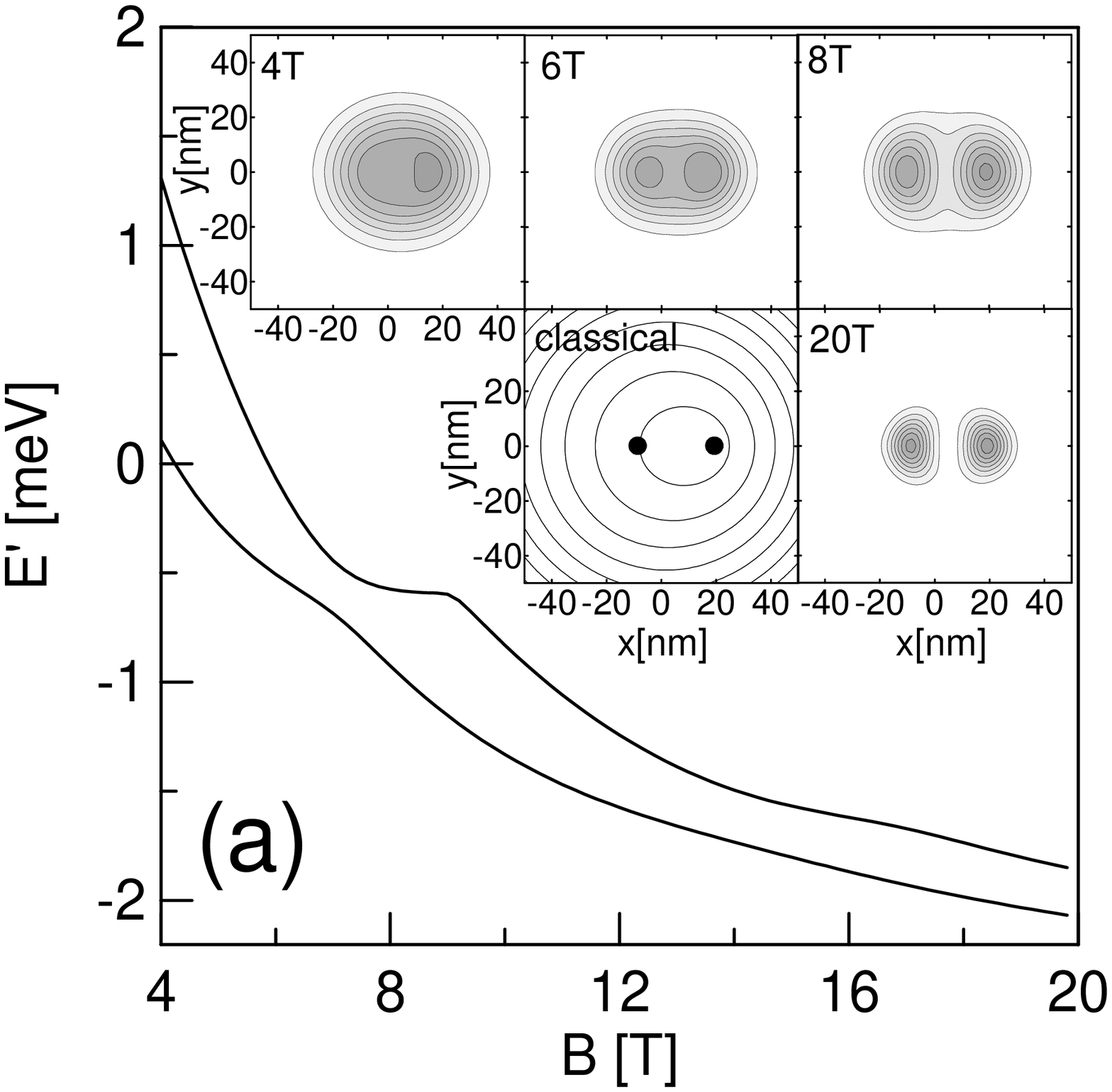}{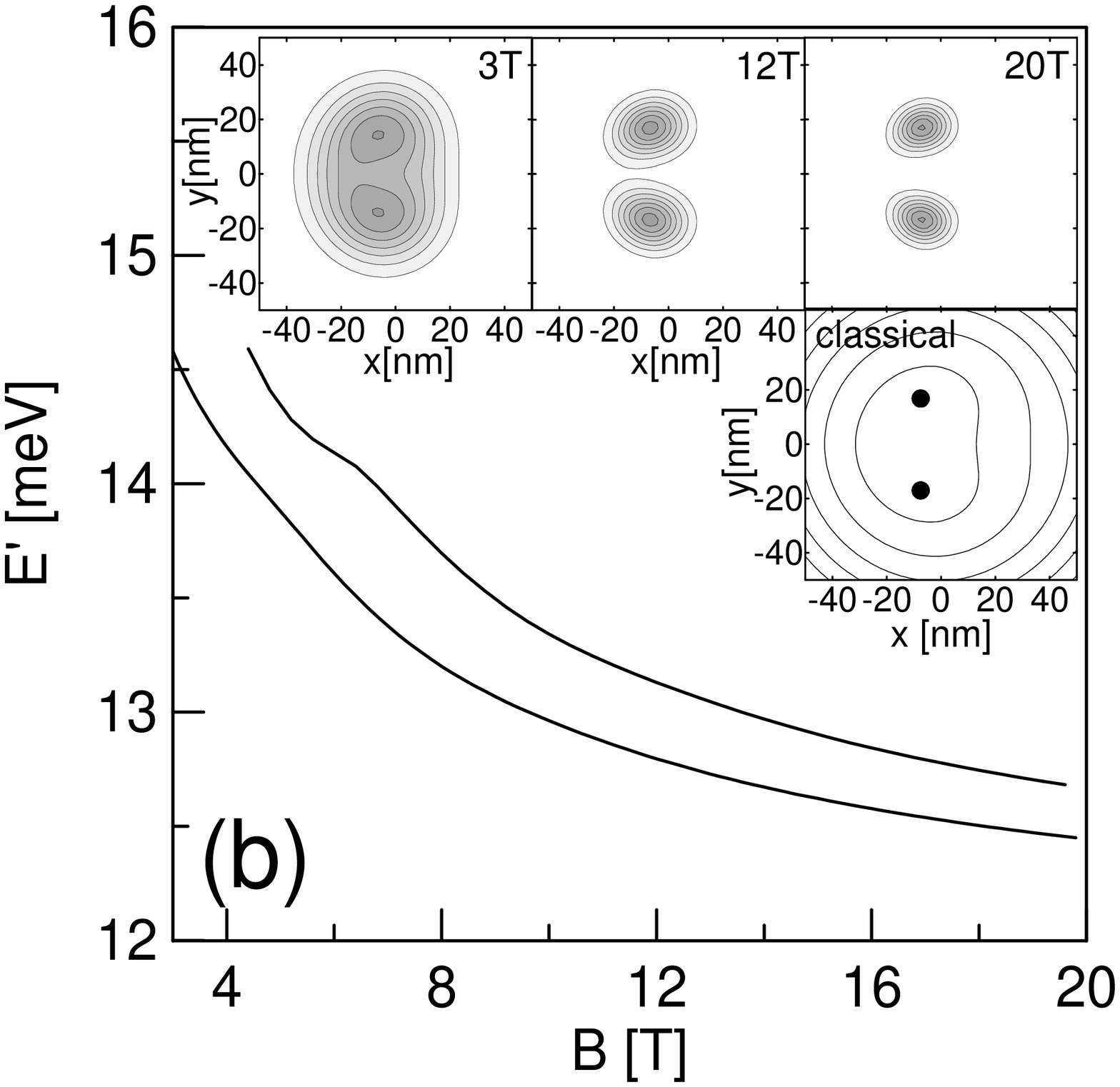}
\caption{Two lowest-energy levels of the 2-electron system in a QD
perturbed by (a)-positively and (b)-negatively charged impurity
situated at (20,0,20) nm. Insets show the ground state charge
densities as well as the classical configurations on a background
of potential profile (the equipotential lines are spaced by 3
meV).
 }
 \label{fig2e20}
\end{figure}

The pinning effect is stronger when the defect is closer to the QD
plane. In the rest of the paper we consider a defect located at
(20,0,20) nm. Fig. \ref{fig2e20} shows the results for 2
electrons. An attractive impurity [fig. \ref{fig2e20}(a)] enhances
the harmonic QD potential which results in a stronger charge
localization and as a consequence shifts the anticrossings to
higher values of the magnetic field. The energy gap between the
lowest levels is larger for repulsive defect [fig.
\ref{fig2e20}(b)]. In both systems an anticrossing related with
the MDD breakdown is visible [$\sim 7$ T in (a) and $\sim 5$ T in
(b)]. Both systems present smooth non-oscillatory convergence to
the lowest-energy configuration of their classical counterparts.

\begin{figure}[htbp]
                \twoimages[width=6.5cm]{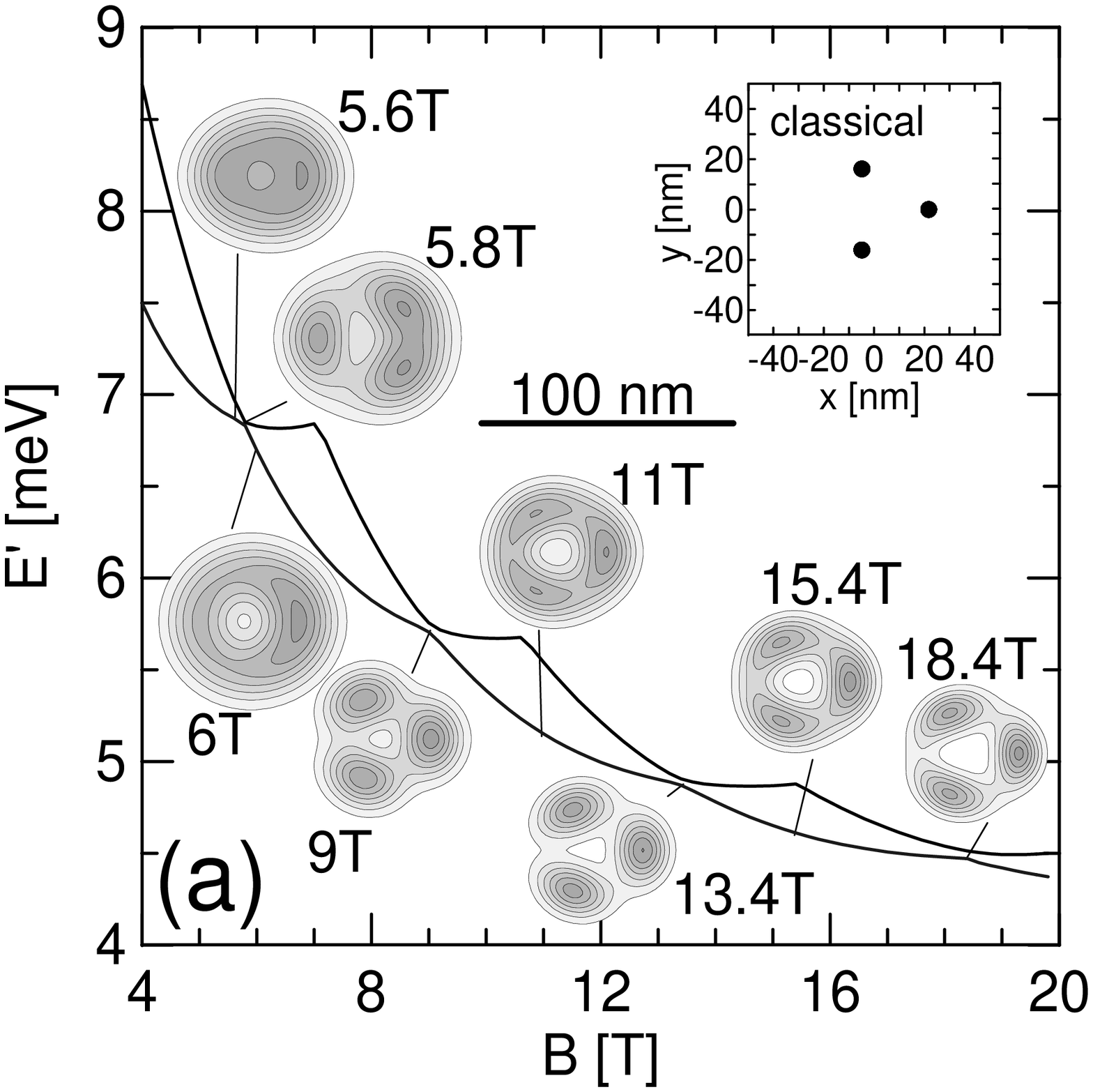}{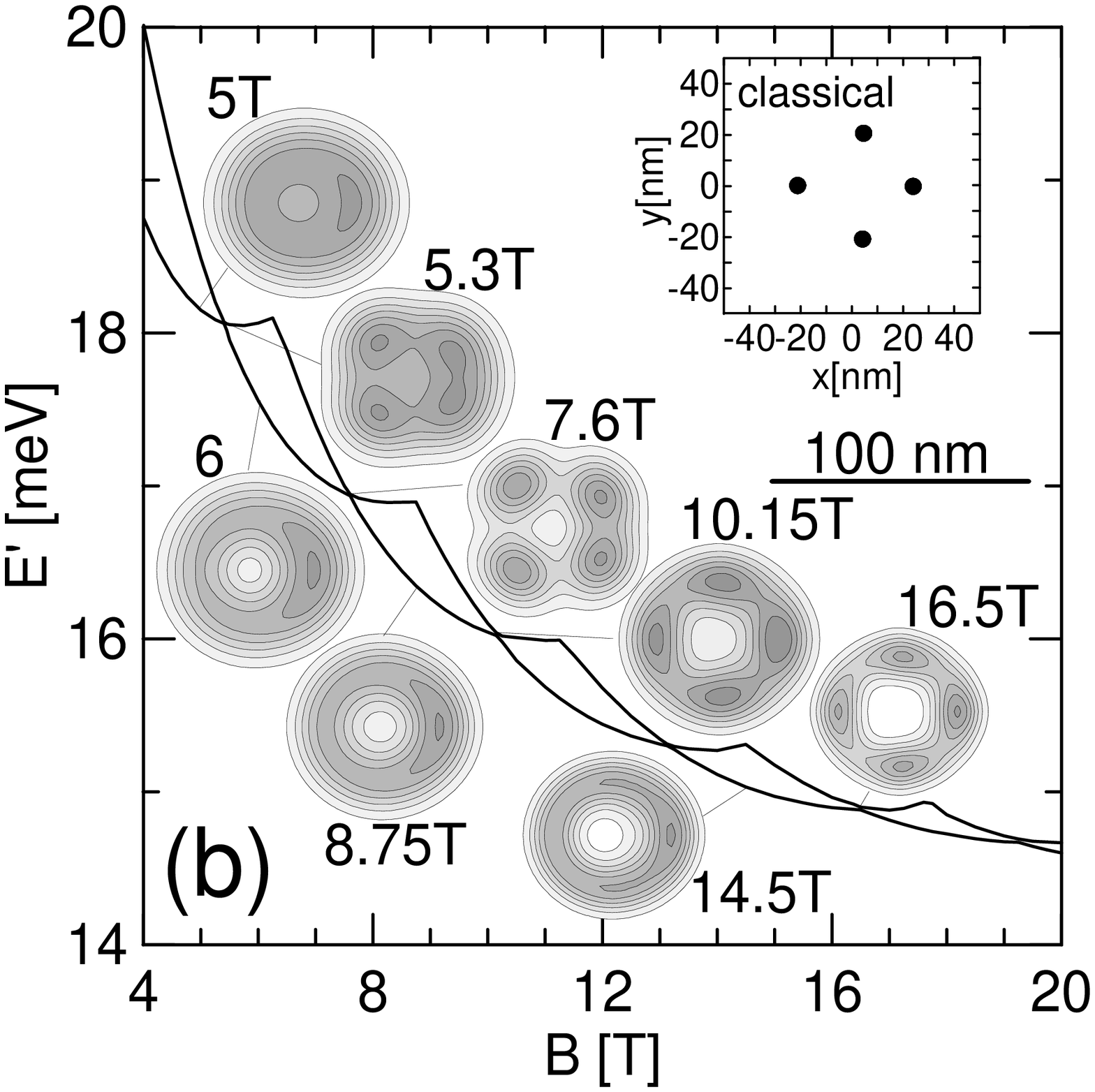}
\caption{(a) Two lowest-energy levels of the 3-electron system in
a QD perturbed by a positively charged defect situated at
(20,0,20) nm. Insets show the ground state charge densities as
well as the classical lowest-energy configuration. (b) Same as (a)
but for 4 electrons.
 }
 \label{fig3e20}
\end{figure}

Fig. \ref{fig3e20} shows the plots for an attractive impurity with
$N=3$ (a) and $N=4$ (b). For both $N=3$ and 4 the energy gaps
between the anticrossing levels remain small [around 0.01 meV
(0.04 meV) for $N=4$ (3)] and the pinning of the WMs exhibits anew
the oscillatory dependence on the magnetic field. The distribution
of charge maxima in the WMs pinned at the MDD breakdown (5.8 T for
$N=3$ and 5.3 T for $N=4$) differs from their classical
counterparts. In classical systems a single electron is trapped
under the attractive impurity. In the WM pinned at the MDD
breakdown 2 electrons fit in the local minimum of the potential
induced by the defect. At higher fields (9 T for $N=3$ and 10.15 T
for $N=4$) the pinning fixes the charge maxima near the
equilibrium positions of classical electrons. Thus an interesting
rotation of the pinned WM is found as function of the magnetic
field. The change in the charge distribution in the WMs between
the MDD decay and the classical limit is similar to the
magnetic-field-induced transformations of the WMs in circular dots
for larger $N$~\cite{MCHF}. At high magnetic field the 3-electron
charge density acquires the semi-classical charge distribution
even between the anticrossings [cf. plots for 11 and 15.4 T in
fig. \ref{fig3e20} (a)]. This is not observed for $N=4$ in the
studied magnetic field range.

\begin{figure}[htbp]
                \twoimages[width=6.5cm]{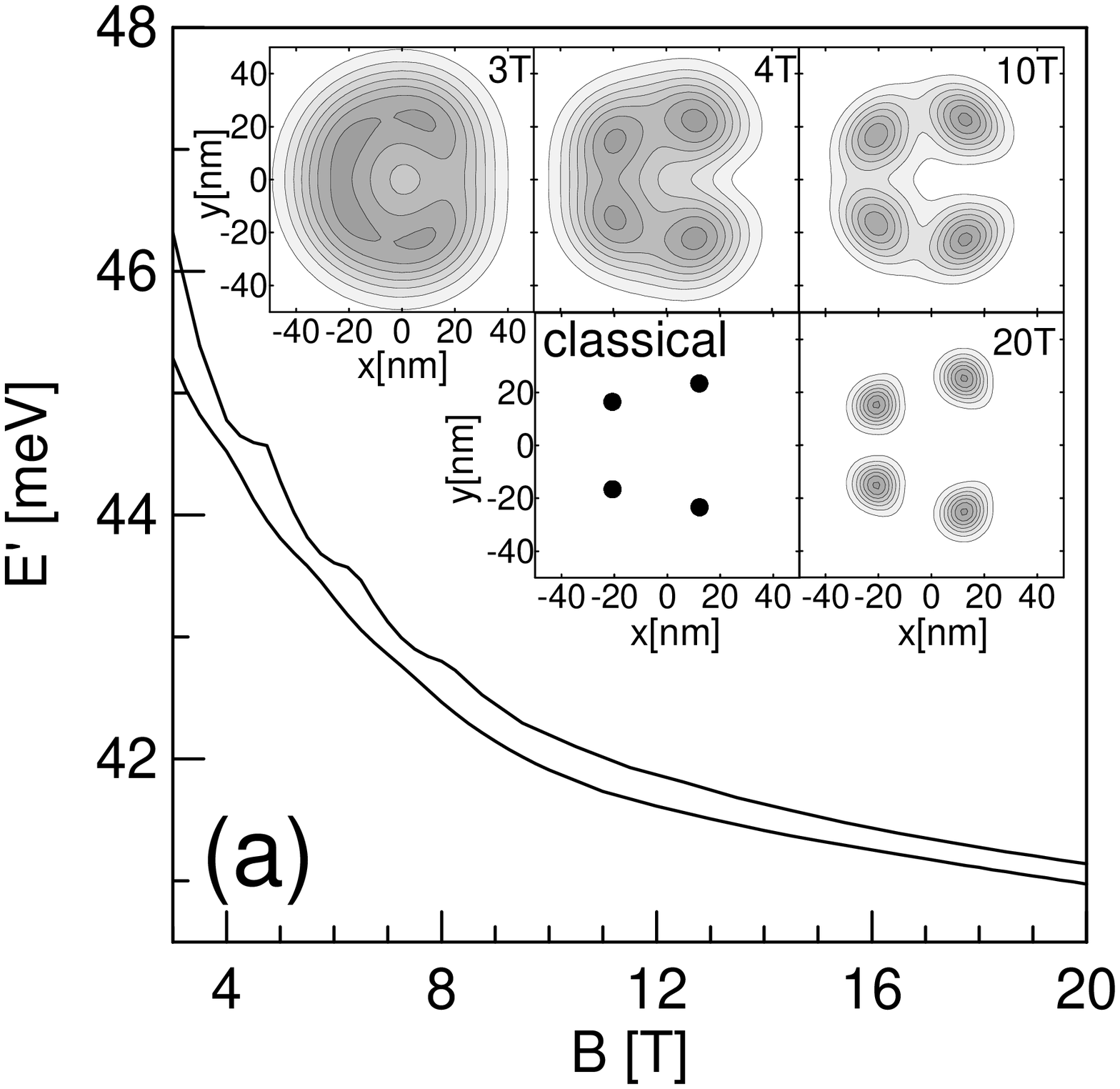}{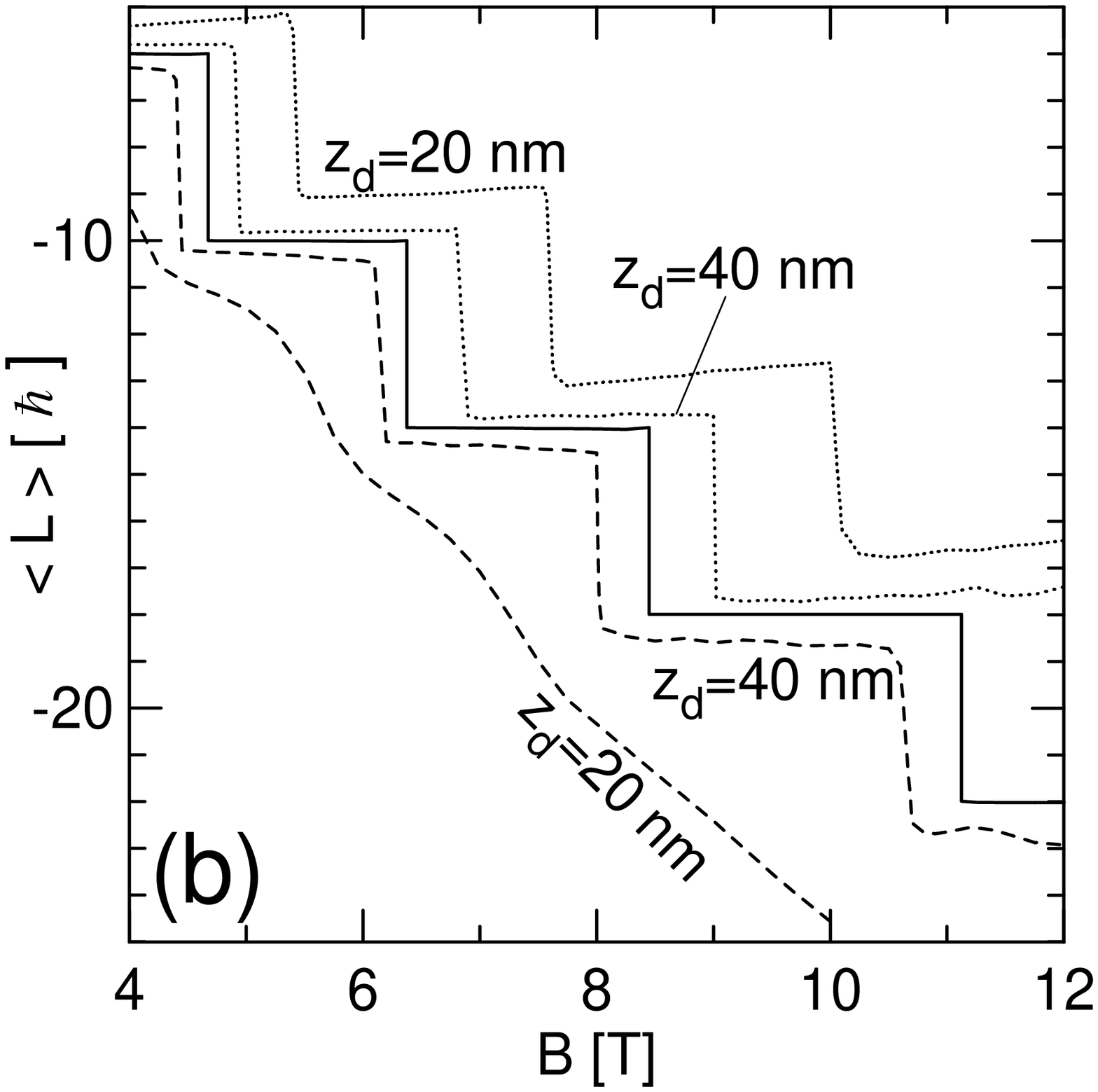}
\caption{(a) Same as Fig. 4 (b) but for a negatively charged
defect. (b) Average value of the total angular momentum for 4
electrons without the defect (solid line), in the presence of an
attractive (dotted lines) and repulsive (dashed lines) defects at
(20nm,0,$z_d$)
 }
 \label{fig5}
\end{figure}

The results for 4 electrons in the presence of a repulsive defect
are shown in Fig. 5. The ground state energy is a smooth function
of the magnetic field and oscillations appear only in the excited
state. A continuous MDD decay appears around 4 T. The charge
density tends in a non-oscillatory way to the classical limit of
point charges.

The influence of the charged defects on the average value of the
total angular momentum for 4 electrons is shown in Fig. 5(b). In
the presence of a defect the average values of angular momentum
take non-integer values and their dependence on the magnetic field
becomes continuous, however much of the stepwise character of a
pure QD is conserved for the positive impurity as well as for a
distant negative deffect. For the positive (negative) defect the
electrons become localized closer (further) from the origin which
results in a decrease (increase) of the absolute value of the
angular momentum with respect to the unperturbed case. For a
negative deffect closer to QD plane the average value is a
smoothly decreasing function of the magnetic field. This fast
increase of the absolute value of angular momentum is related to
the localization of the charge density near the classical
equilibrium points [cf. Fig. 5]. Results for 3 electrons for this
position of the negative defect are qualitatively the same as for
4 electrons.

Comparing the results for an attractive with those of a repulsive
defect [cf. figs \ref{fig3e20} and \ref{fig5}] shows that the
pinning is much more effective in case of a repulsive defect. The
attractive defect enhances the confinement potential of the QD,
decreases its size and hinders the Wigner crystallization itself.
Moreover, it binds one of the electrons in its neighborhood. The
potential of the bound electron and the defect potential partially
cancels and as a consequence the other electrons see a nearly
circular potential and the system in the external magnetic field
behaves essentially like a $N-1$ electron system. On the other
hand, the potential of the repulsive defect is not screened, so it
breaks the circular symmetry of the potential felt by each of the
electrons in a more pronounced manner.

In conclusion, magnetic-field induced WMs in circular dots are
from their very nature susceptible to pinning by the potential of
an external charged defect at the angular momentum transitions.
Our results can be summarized as follows 1) At large distance
between the QD plane and the defect the pinning has a re-entrant
character, i.e., it appears only at the energy level
anticrossings, which are situated near the angular momentum
transition fields of the unperturbed system. 2) For an impurity
placed closer to the QD plane, the pinning by the repulsive defect
is more effective and leads to a non-oscillatory convergence of
the charge density to the classical limit at high field for all
$N$. The pinning effect of a positively charged defect is strong
only for two electrons. For larger numbers of electrons it is
weakened by a partial screening of the defect potential by an
electron trapped in the defects neighborhood so that the
re-entrant pinning behaviour is conserved. 3) For a positively
charged defect close to the QD a magnetic field induced
re-orientation of the WM is predicted.

\acknowledgments This paper has been supported by the Polish
Ministry of Scientific Research and Information Technology in the
framework of the solicited grant PBZ-MIN-008/P03/2003, the Flemish
Science Foundation (FWO-Vl), the Belgian Science Policy and the
University of Antwerpen (VIS and GOA). One of us (BS) is supported
by the Foundation for Polish Science (FNP).


\begin{thebibliography}{0}

\bibitem{WC}
  \Name{Wigner F.P.}
  \REVIEW{Phys. Rev.}{46}{1934}{1002}.


\bibitem{Egger} \Name {Egger R., H\"ausler W., Mak C.H. \and Grabert H.}
\REVIEW{Phys. Rev. Lett.} {82}{1999}{3320}.

\bibitem{Koonin} \Name{M\"uller H.-M. \and Koonin S.E.}
\REVIEW{Phys. Rev. B} {54}{1996}{14532}.

\bibitem{REIMAN} \Name {Reimann S.M. \and Manninen M.}
\REVIEW{Rev. Mod. Phys.} {74}{2003}{1283}.

\bibitem{MAKSYM}\Name{Maksym P.A., Immamura H., Mallon. G.P. \and Aoki H.}
\REVIEW{J. Phys. Condens. Matter} {12} {2000} {R299}.

\bibitem{pinedWC} \Name{Pudalov V.M., D'Orio M., Kravchenko S.V.
\and Campbell J.W.} \REVIEW{Phys. Rev. Lett.} {70}{1993}{1866}.

\bibitem{CDM} \Name{Vdovin E.E. ,Levin A., Patan\`e A., Eaves L., Main P.C., Khanin N.Yu., Dubrovkii Yu.V., Henini M.
\and Hill G.} \REVIEW{Science} {290} {2000}{122}.

\bibitem{Maniani} \Name{Manninen M., Koskinen M., Reimann S.M., \and Mottelson B.}
\REVIEW{Eur. Phys. J. D.}{16}{2001}{381}.

\bibitem{REGGER} \Name{Reusch B. \and Egger R.} \REVIEW{Europhys.
Lett} {64} {2003} {84}.

\bibitem{Bedanov} \Name{Bedanov V.M. \and Peeters F.M.}
\REVIEW{Phys. Rev. B} {49}{1994}{2667}.



\bibitem{magic} \Name{Maksym P.A.} \REVIEW{Phys. Rev. B}
{53}{1996}{10871}.



\bibitem{VQD2} \Name{Oosterkamp T.H., Janssen J.W., Kouwenhoven L.P.,
Austing D.G., Honda T.\and Tarucha S.} \REVIEW{Phys. Rev. Lett.}
{82}{1999}{2931}.

\bibitem{BSA} \Name{Bednarek S., Szafran B.\and Adamowski J.} \REVIEW{Phys. Rev. B}
{64}{2001}{195303}.

\bibitem{sed} \Name{Tavernier M.B., Anisimovas E., Peeters F.M.,
Szafran B., Adamowski J., \and Bednarek S} \REVIEW{Phys. Rev. B}
{69} {2003} {205305}.

\bibitem{Kainz} \Name{Kainz J., Mikhailov S.A., Wensauer A., \and
R\"ossler U.} \REVIEW{Phys. Rev. B} {65}{2002}{115305}.

\bibitem{MCHF} \Name {Szafran B., Bednarek S., \and Adamowski J.}
\REVIEW{Phys. Rev. B}{67}{2003}{045311}.

\bibitem{Yannouleas}\Name{Yannouleas C. \and Landman U.}
\REVIEW{Phys. Rev. B} {68} {2003} {035326}.

\end{thebibliography}
\end{document}